\RequirePackage{fix-cm}
\documentclass[twocolumn]{svjour3}          
\smartqed  
\usepackage{graphicx}
\usepackage{caption}

\newsavebox{\smlmat}
\savebox{\smlmat}{$W=\left[\begin{array}{rrr} 1&-0.4 & 0 \\
 -0.4& 1 & 0 \\
 -1 & -1 & 1
 \end{array}\right]$}

\usepackage{amsmath}
\begin{document}

\title{Data-driven modeling of the olfactory neural codes and their dynamics in the insect antennal lobe}

\subtitle{}


\author{Eli Shlizerman \and Jeffrey A. Riffell \and
        J. Nathan Kutz 
}


\institute{Department of Applied Mathematics
           \and
           Department of Biology \and 
           Department of Applied Mathematics at the University of Washington, Seattle, WA, USA       
}

\date{Received: date / Accepted: date}

\maketitle

\begin{abstract}
Recordings from neurons in the insects' olfactory primary processing center,
the antennal lobe (AL), reveal that the AL is able to process the input
from chemical
receptors into distinct neural activity patterns, called olfactory neural codes.
These exciting results show the importance of neural codes and their
relation to perception. The next challenge is to \emph{model the dynamics}
of neural codes. In our study, we perform multichannel recordings from the
projection neurons in the AL driven by different odorants. We then derive
a neural network from the electrophysiological data. The network
consists of lateral-inhibitory neurons and excitatory neurons, and is capable
of producing unique olfactory neural codes for the tested odorants. Specifically,
we (i) design a projection, an odor space, for the neural recording
from the AL,  which discriminates between distinct odorants trajectories (ii) characterize scent 
recognition, i.e., decision-making based on olfactory signals and (iii) infer the wiring 
of the neural circuit, the connectome of the AL. We show that the constructed model 
is consistent with biological observations, such as contrast enhancement and robustness
to noise. The study answers a key biological question in identifying how
lateral inhibitory neurons can be wired to excitatory neurons to permit
robust activity patterns.

\end{abstract}

\section{Introduction}
\label{intro}
\label{intro}
The analysis of neural recordings from diverse sensory systems shows
that meaningful sensory input is encoded into patterns of
spatial-temporal activity of neuron populations. These populations,
and their corresponding patterns, are termed \emph{cell assemblies}
and have been suggested to serve as the basis for the functional
organization of cells for performing high-level
tasks~\cite{Harris2005,Buzsaki2010}.  The functionality of cell
assemblies is substantiated by their ability to produce unique
patterns of  spatio-temporal activity to encode sensory information
for a given modality~\cite{Mazor2005,HarrisHear2011}. An analog for
such a phenomena is an \emph{encoding} scheme, in which the correct
stimulus produces a robust, and repeatable \emph{neural code} of
spatio-temporal activity.  Evidence of encoding strategies are found
in different sensory systems.  In olfactory and auditory systems,
neural codes take the form of spatial firing-rate (FR) patterns that
are exhibited by the output
neurons~\cite{Stopfer1999,Laurent1999,Laurent2002,Riffell2009}.  The
spatio-temporal encoding patterns were established by the application
of standard data analysis techniques,  e.g., Principal Components
Analysis (PCA), to the time series of FR responses of the cell
assemblies \cite{HarrisHear2011,Broome2006}.  The success of these
methods indicates that the response of cell assemblies is indeed
\emph{low-dimensional} so that for each individual stimulus a unique trajectory in a low-dimensional subspace is
identified.   

With these discoveries, it is intriguing to understand how sensory 
neural networks are designed in order to produce such
behavior.   Specifically, why do the encoding dynamics appear  to be
robust even for noisy stimuli? How is the network structure producing
robust patterns of neural activity?  Obtaining answers  to these 
questions is the next challenge in unraveling the principles of sensory
processing~\cite{Nagel2011,Wilson2008}.   For the first question,
investigations suggest that cell assemblies maintain several
mechanisms for shaping the correct output response. One such mechanism
is known to be \emph{lateral inhibition} \cite{Laurent1999,Egger2003},
where both inhibitory and excitatory neurons receive common input and
interact to mediate the response of excitatory neurons. A hallmark of
lateral inhibition is  an increase in the signal-to-noise ratio in
encoding the  input,  called \emph{contrast enhancement}. The
signature of contrast enhancement is  an increase in the amplitude or
frequency of the response to such that it
can be easily distinguished from the response to random
stimuli~\cite{Yokoi1995,Laughlin1989,Olsen2008,Cook1998,Wilson2008}. 
Experimental observations of contrast enhancement show that the FR representation of the data with the right
dimensionality reduction technique can potentially reveal the sensory
processing mechanisms~\cite{Broome2006}.  
%
%
\begin{figure*}[!t]
\includegraphics[width=1\textwidth]{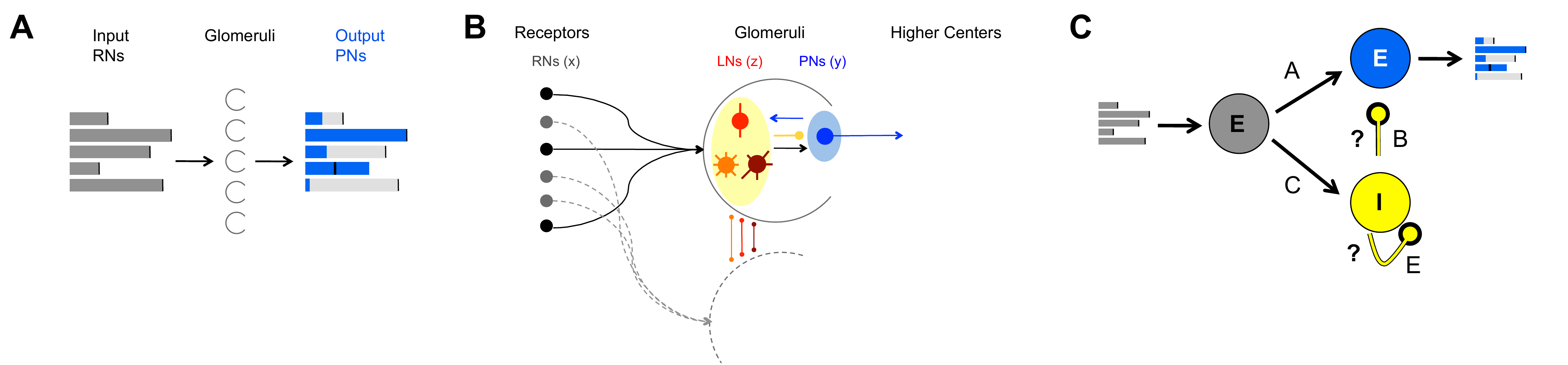}
\caption{ {\bf AL network structure and function.} (A) Demonstration of
a neural code: FRs of the input neurons (gray bars) are processed by neural
dynamics in the Glomeruli (half circles) in the AL  and result with ``shaped"
FRs of the output neurons (blue
bars). (B) Anatomical structure of neuron types and wiring in the AL. Both
PNs (blue shaded) and LNs (yellow shaded) receive input from RNs (black
balls). LNs  synapse to LNs 
and  PNs in other glomeruli via  local (red), global homogeneous (orange)
/ heterogeneous (brown) with mainly inhibitory synapses. PNs as output
neurons have excitatory synapses  to neurons
outside the AL, the mushroom body.   (C) Schematics of a network that mimics
the wiring in the AL in
moths divided into three populations: input excitatory RNs (gray), interneurons
inhibitory LNs (yellow) and   projection excitatory PNs (blue). A,B,C and
E denote the connectivity matrices between different populations of neurons,
i.e.,
the connectome.
With correct calibration of the inhibitory connections, marked by the question
marks, the network  can produce  neural codes as in (A). }
\label{fig:1}
 \end{figure*}  

For the question of determining the network structure that is producing the 
robust neural codes it is necessary to model  the actual network of neurons
responsible for the encoding. The modeling procedure involves reconstruction of the
\textit{network wiring}, i.e., modeling individual neuron dynamics and
their network interactions \emph{(connectome)}\cite{Seung2011,Jbabdi2012}. The
connectome of different sensory neuronal networks may vary. For
example, in vision, the retinal ganglion cells are ordered such that
locally neighboring cells are responsive to neighboring parts of the
visual stimulus, termed a {\em retinotopic map}~\cite{Bock2011}. In
olfaction, output neurons are also selective for certain odorant
stimuli, providing a {\em chemotopic map}.  However, the neighboring
output neurons are not necessarily similar in their tuning to specific
chemicals. Instead  lateral inhibition mediates and shapes the
responses of the output neurons, resulting in an \emph{effective
chemotopic map} between the input  and the output cells 
\cite{Linster2005,Reisenman2008,Cleland2005}. Resolving this  mapping
is critical for determining how neurons process chemical information. 

To both unravel the mechanisms responsible for the observed
low-dimensional phenomena, and easily test the model for consistency
with  experimental dynamics, the neuronal network  should  be ideally
modeled with a dimensionally reduced \emph{dynamical} system. The
derivation of such a system can be accomplished with two possible
approaches.  A bottom-up approach would model the network at the
detailed synaptic level with
many neurons composing a network and each neuron being a nontrivial
dynamical system~\cite{Buckley2011,Christie2006}. However, the complexity
of  the high-dimensional
system makes the extraction of a low-dimensional model
 challenging. Additionally, for many systems there
is not enough anatomical and physiological data available to calibrate
such a high-dimensional network.  A top-down approach, on the other
hand, would attempt to construct a low -dimensional model from the
available  anatomical evidence, fitting key parameters based upon
experimental observations.  The model can then be explored with the
aim of creating similar patterns to those observed in the actual
system~\cite{Afraimovich2004,Rabinovich2001,Abbott2007,Luo2010,Brody2003}. The primary drawback of this approach is that a general
framework for calibration of the model to experimental data does not
exist~\cite{Mazor2005,Rabinovich2008,Buckley2012}.

To overcome this difficulty, we  introduce the
\emph{dynamical dimension reduction} method that takes the top-down approach
in conjunction with experimental data. The outcome of the approach is
a high-dimensional system that exhibits  low-dimensional dynamics. The
method is fundamentally different than the standard top-down approach,
as it does not  determine parameters by  simulation and fitting.
Instead, it  projects a high-dimensional dynamical system, using
\emph{proper orthogonal decomposition}, onto orthogonal  modes. It
then matches the projected dynamical system with the conjectured
low-dimensional dynamical system. Plugging in the experimental  modes
as orthogonal modes establishes the full network wiring. The
constructed model  can be then easily compared with the experimental
dynamics for consistency. With this methodology we  model the
olfactory processing unit, the \emph{antennal lobe} (AL) in the {\em
Manduca sexta} moth,  a well-characterized physiological and
behavioral experimental neural system in olfaction
\cite{Riffell2009,Reisenman2008}. The experimental data utilized are  
the observed neural codes, which are an efficient representation of the
measured spatio-temporal neural responses. The underlying concept of
our  method is that the dynamics projected onto the neural codes
exhibit a trajectory towards a fixed point for each input key, a
feature consistent with experimental observations. From the
low-dimensional dynamics we are able to infer the  network wiring for the
AL. The construction of such a
computational model is thereby  able to answer  both questions posed
above since it establishes (i) a dynamical system for the FR pattern
and (ii) a connectome that mimics the connectivity responsible for
robust neural codes.

\section{Methods and Models}
\label{sec:methods}
\subsection{Data Driven Top-Down Modeling Approach}
\begin{figure}[!t]
\centering
\includegraphics[width=0.5\textwidth]{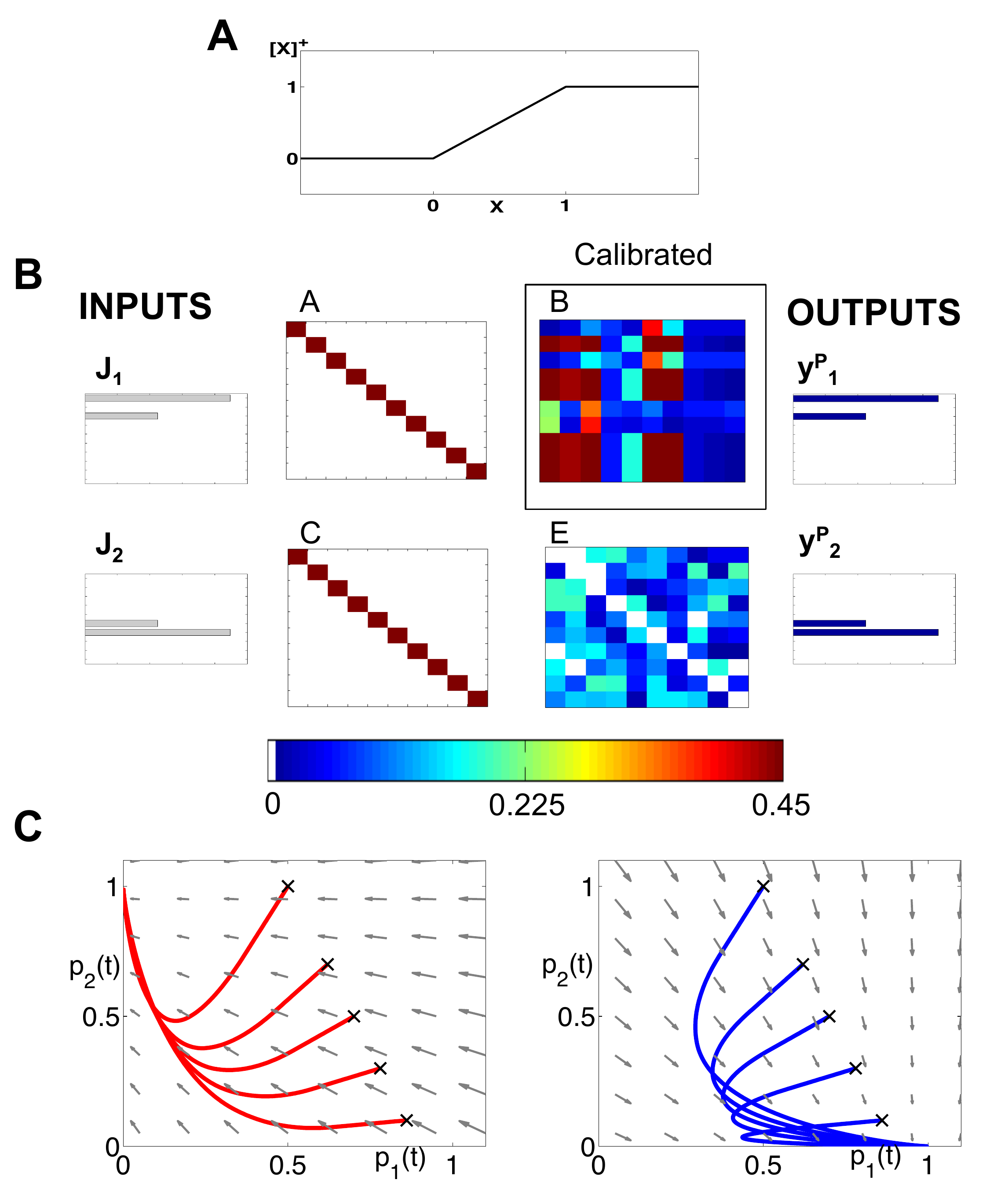}
\caption{{\bf Modeling approach to the dynamics of the AL.} (A) Piecewise linear f-I  curve that we use to model the neurons as FR units. (B) Example of
a calibration of a network, consisting of ten neurons
from each class and encodes two neural codes. The input keys $\vec J_1,\vec
J_2$,
and the population vectors $\vec y_1^P, \vec y_2^P$ with given connectivity matrices $A,C,E$ allow
us to reconstruct the matrix $B$. We use the matrix~\usebox{\smlmat} 
 to prescribe the  low-dimensional projected system. (C)
 Deterministic
dynamics when
 the input is the  key: $\vec J_1$ (left) or $\vec J_2$ (right). 
 When $\vec J_1$
is applied, the fixed point is located at $(p_1,p_2)=(1,0)$ and all trajectories
 are attracted to it (see  blue sample trajectories with initial conditions denoted by `x'). When $\vec J_2$ input
is applied, the trajectories are being attracted to $(p_1,p_2)=(0,1)$ (see
red trajectories).}
\label{fig:2}
\end{figure}

The neural cell assemblies participating in the processing of olfactory
information in the AL are the receptor cells (RNs) that carry the input
from the environment, the projection (output) neurons (PNs), and local
interneurons (LNs) (reviewed by \cite{Hildebrand1997,Hansson2000,Martin2011}).
As in the anatomical grouping, we model the network as a set of
 three firing-rate units corresponding to the interacting cells in the
 olfactory AL:
\begin{eqnarray}
\label{eq:FRsys1}&&\dot  {\vec x} = - \vec x +\vec J, \\
\label{eq:FRsys2}&&\dot{\vec y} = -\beta \vec y + [A\vec x - B \vec z]^+, \\
\label{eq:FRsys3}&&\dot{\vec z} = -\gamma \vec z + [C\vec x - E\vec z]^+.
\end{eqnarray}
The three groups (vectors) $\vec x, \vec y$ and $ \vec z$ in these equations correspond to the three anatomical groups  RNs,  PNs 
and LNs respectively. The input into the PNs and LNs  is modulated by a standard linear threshold function
denoted by $[.]^+$, as in~\cite{Linster2005}.  Figure~\ref{fig:2}A  illustrates the threshold function used here. The vector
$\vec J$ is the external input into the receptor cells which is driven by
the chemosensory processes in the antenna.

In the deterministic version of this model, where the input is either constant
or   time dependant, the dynamics can be intuitively described.
Specifically, when there is significant input into the population of receptor neurons
  ($\vec x$), these neurons lock onto the driving input $\vec J$~\cite{Buckley2011,Buckley2012}.
In the case of constant input, the receptor population will converge to
a fixed point $\vec x_0= \vec J$. This in turn excites both the projection neurons ($\vec y$) and
the interneuron populations ($\vec z$). A \emph{meaningful} input should
 excite
a  spatial \emph{stable}  pattern $\vec y^{P}$  in the projection neurons.
Here, the stable
spatial patterns $\vec y^{P}$ are thought of as library elements which distinguish (encode)
various recognized odorants.
Note that the pattern is not necessarily equal to the input, i.e.,
$\vec y^P \neq \vec J$.  

Our goal is to understand how  the network in Eqs.~(\ref{eq:FRsys1}-\ref{eq:FRsys3}) 
can be made capable to produce stable patterns and discriminate between them. Particularly, we would like to find a network connectome, consisting
of the  connectivity matrices $A,B,C,$ and $E$, that enhances the components
in the input that correspond to recognized patterns ($\vec y^{P}$) and inhibit other remaining
components. In practice, the structure of the connectivity matrices $A$ and $C$ is local and can be 
obtained  from  anatomical experimental knowledge, while the  structure of the matrices $B$ and $E$ is  mostly unknown. 

The method   \textit{dynamical
dimension reduction}      that we introduce in this work provides a procedure  to construct the unknown matrices $B$ and $E$. 
The first step in the method   is to obtain population encoding
vectors (orthogonal patterns
  $\vec y^{P}$) from the electrophysiological recordings of PN neurons. We then show that a  projection of the PN dynamical equations, Eq.~(\ref{eq:FRsys2}),
onto the population encoding
vectors provides a division of these equations into two models: a reduced model for the dynamics of population vectors
and  a model for the dynamics of  remaining patterns.  Separating the
system into two  models
allows us to impose constraints on the dynamics of each model. Particularly,
we require  stable patterns in the reduced model and rapid decay of the remainder. We show that these requirements form a convex minimization problem
which solution is the unknown connectome.


The projection is done as follows.  If the system does not saturate, then the excitable
regime can be modeled by a linear version of  Eqs.~(\ref{eq:FRsys2})-(\ref{eq:FRsys3})
in which the brackets from the saturation terms are  removed. Additionally, if the $\vec x$ dynamics are fast in comparison to those of $\vec y$ and $\vec z$ 
(RNs drive the response in LNs and PNs), then
$\vec x$ can be replaced by the input $\vec J$, i.e. its fixed point, and we derive the following
system
\begin{eqnarray}
\label{eq:FRsys2lin}&\frac{d \vec y}{dt} = -\beta \vec y + A\vec J - B \vec z, \\
\label{eq:FRsys3lin}&\frac{d \vec z}{dt} = -\gamma \vec z+ C\vec J- E\vec z.
\end{eqnarray}
In this system, the vector $\vec y(t)$ describes the  dynamics of the coefficients
of a standard basis ($y_i(t)$ is the dynamics of $i$-th PN neuron). However,
we are interested in determining the dynamics of the observed patterns. From
this representation, it is not immediately clear how to conclude that coding patterns
in $\vec y$ appear while others do not, and what kind of connectivity
matrices support such formations.  Thus the next step in our analysis is
to  decompose the system into  the encoding patterns
and  a remainder. For such a decomposition, we assume that there
is a library matrix $L$ of observed patterns $L=\{\vec y_1^{P},...,\vec
y_l^P\}$. We take into account that the library is a semi-positive matrix and we normalize
each column vector (pattern) in the matrix. We then transform the matrix to an \textit{orthonormal matrix} $O^P$.
In this matrix, each column vector  is called  a \textit{population encoding vector}
and represents  neurons
and their expected  firing-rates evoked by a particular input-key. The transformation to the orthonormal matrix
is achieved  by applying a threshold and a maximum rule
on each element $l_{ij}$ of the matrix $L$. Thereby each element $o^P_{ij}$
in the matrix $O^P$ is defined as follows
 \begin{eqnarray*}o^P_{ij}=U_1(l_{ij})=\left\{
        \begin{array}{ll}
                l_{ij}  & \mbox{if } l_{ij}=\max{ (\vec l_i)} \geq \tau. \\
                0 &\text{otherwise}
        \end{array}
\right.
 \end{eqnarray*}
where $\tau$ is the threshold value (chosen as $\tau=0.07$ in Fig.~\ref{fig:4}). This construction results in  a matrix with a single positive element in
each row vector or a zero row vector, i.e., the system is effectively made orthogonal. The zero row vectors  indicate PN
neurons that do not substantially contribute to any of the patterns and
thus these neurons will be considered to belong to the \textit{remainder} vector. To construct the remainder vector, $\vec
o^R$, we define the transformation 
$U_2$ 
 \begin{eqnarray*}\vec o^R =U_2(U_1(l_{ij}))=\left\{
        \begin{array}{ll}
                1  & \mbox{if } \max{ (\vec l_i)} =0 \\
                0 & \text{otherwise}.
        \end{array}
\right.
 \end{eqnarray*}
that assigns the value of unity  if   the corresponding row in $O^P$ that
is a zero vector. As a final step we normalize  $\vec o^R$ and augment the matrix $O^P$ with the vector $\vec o^R$
to create the matrix $O$:
\begin{eqnarray*}
&L=\left[\begin{array}{ccc}
\vdots &  & \vdots \\
\vec y_1^{P} & \hdots & \vec y_l^{P} \\
\vdots &  & \vdots \\
\end{array}\right]_{N\times l}
\to^{U}
O=\left[\begin{array}{cccc}
\vdots &  & \vdots & \vdots\\
\vec o_1^{P} & \hdots & \vec o_l^{P}& \vec o^{R} \\
\vdots &  & \vdots & \vdots \\
\end{array}\right]_{N\times l+1}.
\end{eqnarray*}
  This allows us to describe the dynamics of PNs  with the following
low rank decomposition 
\begin{eqnarray}
\label{eq:decomp} 
\nonumber \vec y(t) &= p_1(t)\vec o_1^{P}  +...+ p_l(t)\vec o_l^{P}+r(t)\vec o^{R}\\
&=O
\left(\begin{array}{c}
p_1(t)\\
\vdots\\
p_l(t)\\
r(t) 
\end{array}\right)
=O \vec p
\end{eqnarray}
Here we multiply each of the population vectors (stationary) by a dynamical coefficient $p_j(t)$
 and the remainder population vector by  $r(t)$.
To derive the equations for the dynamics of the coefficients $\vec p(t)$,
we substitute the decomposition of Eq.~(\ref{eq:decomp}) into Eq.~(\ref{eq:FRsys2lin}) and multiply the equations for $\vec y$ by the transpose matrix $O^{T}$
and use the  fact that for semi-orthogonal matrices  $O^{^T}O=I$.   Thus
 \begin{eqnarray}\label{eq:projeq}
\frac{d \vec p}{dt} &&= -\beta \vec p + O^{T}(A\vec J - B \vec z), \\
\nonumber \frac{d \vec z}{dt}&&= -\gamma \vec z+ C\vec J- E\vec z.
\end{eqnarray}
 This
projection technique is based  on the Proper Orthogonal Decomposition method
introduced in~\cite{Sirovich1987,Sirovich1996} and 
applied to reduction of neuronal networks  in~\cite{eshamee11}. 

In this section we consider the case where the input is 
time-independent and in the Results section explore the system dynamics
with time dependent and noisy inputs.  Since the dynamics in $\vec z$ are independent of the dynamics in $\vec
p$, we can  solve the second equation in Eqs.~(\ref{eq:projeq})
for a fixed point (${d \vec z}/{dt}=0$) 
\begin{eqnarray*}
\vec z_{0}=\tilde{E}^{-1}C\vec J, ~~~~~~\tilde{E} = E+\gamma I  \, .
\end{eqnarray*}
Assuming that the dynamics of LNs are faster
than of PNs, since LNs typically respond with bursting and release fast GABA-A transmitters (within 1-2 ms)~\cite{Christensen1998}, we can plug the expression of the fixed point into the first equation
\begin{eqnarray}
\label{eq:pattdynsys} 
&&\frac{d \vec p}{dt} = \tilde{M} \vec p+ \vec J^{eff}\\ 
\nonumber &&~~~~~~~\tilde{M}=-\beta I,~~\vec J^{eff}= O^{T}(A -B \tilde{E}^{-1}C)\vec J.   
\end{eqnarray} 
The resulting reduced system is a linear nonhomogeneous system of ODEs.
%
%
It has terms that include the parameter  $\vec p$ (multiplied by $\tilde{M}$) and nonhomogeneous terms that are the effective input. Note that since
there  is no input from $\vec y$ into $\vec z$ in
Eqs. (\ref{eq:FRsys2lin})-(\ref{eq:FRsys3lin}),  the homogeneous term  is multiplied by a diagonal matrix $\tilde{M}$. 
The matrix $\tilde{M}$ has only negative eigenvalues ($\lambda_i =-\beta$)
and thus by Lyapunov's stability theorem the model in Eq.~(\ref{eq:pattdynsys}) is globally asymptotically stable, i.e the system will always converge
to a stable equilibrium $\vec p_{0}=(1/\beta) \vec J^{eff}$~\cite{Gajicbook},
see Fig.~\ref{fig:2}C. In systems which
have additional input from the $\vec y$ population into the $\vec z$ population,
the matrix  $\tilde{M}$  will involve non-diagonal terms that express interactions of the patterns. For such wirings it should be verified that the system
is stable, i.e. the dynamics are as in Fig.~\ref{fig:2}C, by solving the Lyapunov equation that will involve the connectivity matrices~\cite{Gajicbook}. 
The solution of the equation, if exists, will impose constraints on the configuration of the connectivity matrices such that the fixed point is stable. 
These constraints will be added to the optimization problem \ref{eq:minprob}. 

While the stability theorem assures that the dynamics of the patterns converge to an equilibrium, it does not guarantee separation of equilibria, which
is required for a robust encoding-decoding system.   Moreover, the {\em
matrices $B$ and $\tilde{E}$
are unknown}, both in theory and in practice.  For that purpose we need to calibrate
the effective input into the population encoding vectors. Following the
same procedure as for the output patterns we construct an \textit{orthogonal
library matrix}, $J_0$, for the input  keys. Then the calibration  is reduced to solving
the following system of underdetermined equations  
\begin{eqnarray}
\label{eq:calequal}
&O^{T}(A  -B \tilde{E}^{-1}C) J_0 = W.
\end{eqnarray}
with the prescribed matrix $W$ of dimensions $(l+1) \times (l+1)$ representing the calibration,
and $B$ and $\tilde{E}$ are the  unknown matrices.
Essentially, this is a linear system of equations with a specified right hand side matrix $W$
where the matrix elements of $W$ determine physiologically relevant characterization of
the importance of various odors.  This is a highly undetermined
set of equations that allows for an infinite number of solutions, i.e. there are
an infinite number of ways to specify $B$ and $E$.  Thus
further constraints must be imposed in order to arrive at a biophysically valid solution.

Each row in $W$ encodes the effect of  the different input keys,
including the remainder,  on
a particular population encoding vector.
For example,   the element on the
$i$-th row
and  $k$-th column, $w_{i,k}$, defines how $\vec J_k$ excites or inhibits
$p_i(t)$. The elements of $W$  are set as follows
\begin{eqnarray}\label{eq:Wdef}
&&
W = 
\left[\begin{array}{cccc}
\ddots& &&\\
\hline
&w_{i,i}&w_{i,k}&w_{i,l+1}\\ \hline
&&\ddots&\\ \hline
w_{l+1,1}&&w_{l+1,k}&w_{l+1,l+1}
\end{array}\right]. 
\end{eqnarray} 
The diagonal element
on the $i$-th row, $w_{i,i}$, defines how $\vec J_i$ affects $p_i(t)$, its
corresponding population encoding vector,
and   has to be set as positive (excitatory).   The input from
the other keys, $\vec J_k$, $k \neq i$, is encoded by $w_{i,k}$ and can be set $0$ or negative. The input from the last key is the input from  the remainder and is encoded by $w_{i,l+1}$. The value
of this element
should be  strictly set to $0$, such that the remainder does not
have  excitatory or inhibitory effect on the population encoding vector.
The last row in $W$ denotes the  input into the remainder
  and thereby the elements, except the diagonal element on that
row should be always negative. See the caption of Fig.~\ref{fig:2} for
a possible configuration of the matrix $W$.

When $A$ and $C$ are  known matrices, then the calibration is accomplished by solving an
inverse problem to find the
matrices $\tilde{E}$ and $B$ that satisfy these equations. Notice that the equations
are underdetermined, i.e., the dimensions of $W$ are
much lower than of $B \tilde{E}^{-1}$,  indicating that the matrices $B$
and $\tilde{E}$ that satisfy  Eq.~(\ref{eq:calequal}) are non-unique.  To find the
appropriate candidates
for  the matrices, we  reformulate
the inverse problem as a convex minimization problem
\begin{eqnarray}\label{eq:minprob}
\text{minimize}~~||O^{T}(A  -B \tilde{E}^{-1}C) J_0 - W||_{Fr}\\
\nonumber \text{subject~to}~~~~B, E\geq\ 0,~~~~~~~~ 
\end{eqnarray}
where $\|\cdot\|$ is the Frobenius matrix norm. When the lateral connections between PNs and LNs are exclusively inhibitory the matrices $B,\tilde{E}$ are  non-negative. When one of the matrices is set to particular wiring
(e.g. $\tilde{E}$ is random) we need to determine only one matrix and the minimization problem is a semi-definite
convex minimization. When there are  excitatory lateral connections or the zero minimum cannot be attained, the semi-definite constraint is relaxed. Another possibility for negative terms in  $B$ and $\tilde{E}$ is when the input keys and the output codes  differ from each other in dimensions.  Indeed, the matrices $B$ and $\tilde E$ permute the lateral effect of the interneurons to support such a coding scheme. Due to many degrees of freedom in the problem, additional constraints
can be added. For example we can  restrict the magnitudes of the elements in $B$ and $\tilde E$ not to exceed a particular value.  Moreover, the
calibration is particular to the choice of the matrices $A$ and $C$ (see
an example in Fig.~\ref{fig:2}B). To solve the minimization problem~(\ref{eq:minprob}) or its variants, we employ the disciplined convex
optimization package CVX implemented in MATLAB~\cite{cvx}.
   
For  input keys being identical to the output
population vectors, i.e. $J_0\equiv O$, 
the calibration creates a system that for a significant magnitude of one
of the input keys, noise and other  population encoding vectors will be
suppressed to allow for a decoding of the input-key (see Fig.~\ref{fig:3}).
Effectively this is a mechanism that produces contrast enhancement as we
discuss in the Results section.

\subsection{Electrophysiological preparation and stimulation}

\emph{Manduca sexta} L. (Lepidoptera: Sphingidae) larvae were obtained from the \emph{Manduca}-rearing facility of the Department of Biology of the University of Washington.  Larvae were reared on artificial diet \cite{Bell76} under long-day light:dark (LD) regimen (LD 17:7) at 25–-26 $^{\circ}$C and 40–--50\% relative humidity (RH), and prepared for experiments 2-–3 d after emergence. In preparation for electrophysiological recording, the moth was secured in a plastic tube with dental wax, leaving the head and antennae exposed. The preparation was oriented so that both ALs faced upward, and the tracheae and sheath overlying one AL were carefully removed with a pair of fine forceps. The brain was superfused slowly with physiological saline solution throughout the experiment.  
        
Electrophysiological recordings were made with 16-channel silicon multielectrode recording arrays (a4$\times$4--3mm50--177; NeuroNexus Technologies, Ann Arbor, MI, USA). This microprobe allows the recording of neurons throughout the AL because of the probes dimensions, with four shanks spaced 125 $\mu$m apart, each with four recording sites 50 $\mu$m apart\cite{Christensen2000,Riffell2009a}.  The probe was positioned under visual control using a stereo microscope. We use routine histological methods (e.g., Riffell et al.~\cite{Riffell2009}) to visualize the tracks left by the probes and identify the recording sites. 
Neural ensemble activity was recorded simultaneously from the 16 channels of the recording array using a RZ2 base station (Tucker-Davis Technologies, Alachua, FL USA) and a PZ2 peamplifier. Spiking data from 16 channels (recorded at four sites on each of the 4 probes)   were extracted from the recorded signals  and digitized at 25 kHz using the Tucker-Davis Technologies data-acquisition software. Spike data were extracted from the recorded signals in the tetrode configuration and digitized at 25 kHz per channel. Filter settings (typically 0.6--3 kHz) and system gains (typically 5,000--20,000) were software adjustable on each channel. Spikes were sorted using a clustering algorithm based on the method of principal components (PCs). Only those clusters that were separated in three dimensional space (PC1--PC3) after statistical verification (multivariate ANOVA; $P < 0.05$) were used for further analysis (6--15 units were isolated per ensemble; $n = 11$ ensembles in as many animals). Each spike in each cluster was time-stamped, and these data were used to create raster plots and to calculate the instantaneous firing-rates (iFRs). Based on the spiking activity, recorded spike trains were identified as an LN or PN     (as in \cite{Lei2011,Riffell2009,Brown2004,Riffell2013}). All analyses were performed with Neuroexplorer (Nex Technologies, Winston-Salem, NC, USA), or MATLAB (The Mathworks, Natick, MA, USA), using a bin width of 5 ms, unless noted otherwise.

Olfactory stimuli were delivered to the preparation by pulses of air from a constant air stream were diverted through a glass syringe containing a piece of filter paper bearing floral odors. The stimulus was pulsed by means of a solenoid-activated valve controlled by the acquisition software (Tucker-Davis Technologies, Alachua, FL USA). AL neurons were stimulated with two odorants: $\beta$-myrcene, a plant-derived odorant used to attract moths \cite{Riffell2009}; and E10,Z12-hexadecadiennal (bombykal \{Bal\}), the primary component of the conspecific female's sex pheromone \cite{Tumlinson1989,Tumlinson94}. Stimulus duration was 200 ms, and five pulses were separated by a 10 s interval. The stimulus durations reflect the time periods in which moths encounter odors when flying in their natural environment \cite{Murlis81,Riffell2008}, and the odorants used to stimulate the preparation are behaviorally effective stimuli, thus allowing neurobiological experimentation in a naturalistic context for discovering how neural circuits process odor information.

\section{Results}

\label{sec:results}
To study the AL's neural encoding dynamics we computationally model the AL as  a  network with each neuron modeled as  a FR unit.
In keeping with the populations of AL cells, three populations of FR units are considered: receptor cells that carry the input from the
periphery (RNs), projection (output) neurons (PNs),  and local
inhibitory interneurons (LNs)  (Fig.~\ref{fig:1}). 
  The dynamics of the populations are  represented by the state vectors,  $\vec x,~ \vec y$ and $\vec z$ corresponding to dynamics of RNs, PNs
and LNs respectively.  Each FR unit in each population  is modeled
by a differential equation that describes unit's self-dynamics (decay in the absence
of input),  interaction with other units  and response to odorant stimulus (for a detailed
description of the construction see the Methods
section).  The network can be calibrated to
perform encoding functions, i.e., produce neural codes. Specifically,  for  each FR pattern that the  PNs
population exhibits  (called \emph{population encoding vector}), there
is a FR pattern
of the RNs population that evokes it (called \emph{input key}) (depicted
in Figs.~\ref{fig:1},\ref{fig:2}).
The results that we obtain from constructing the network   establish how  neurons' connectivity and network dynamics are linked together to produce these encoding functions. Analyzing computational dynamics and comparing
them with experimental
dynamics elucidates what are the typical dynamics  of neural codes and
how they can be perceived. We describe our results in detail below.
%
%
\begin{figure*}[!t]
\centering
\includegraphics[width=0.65\textwidth]{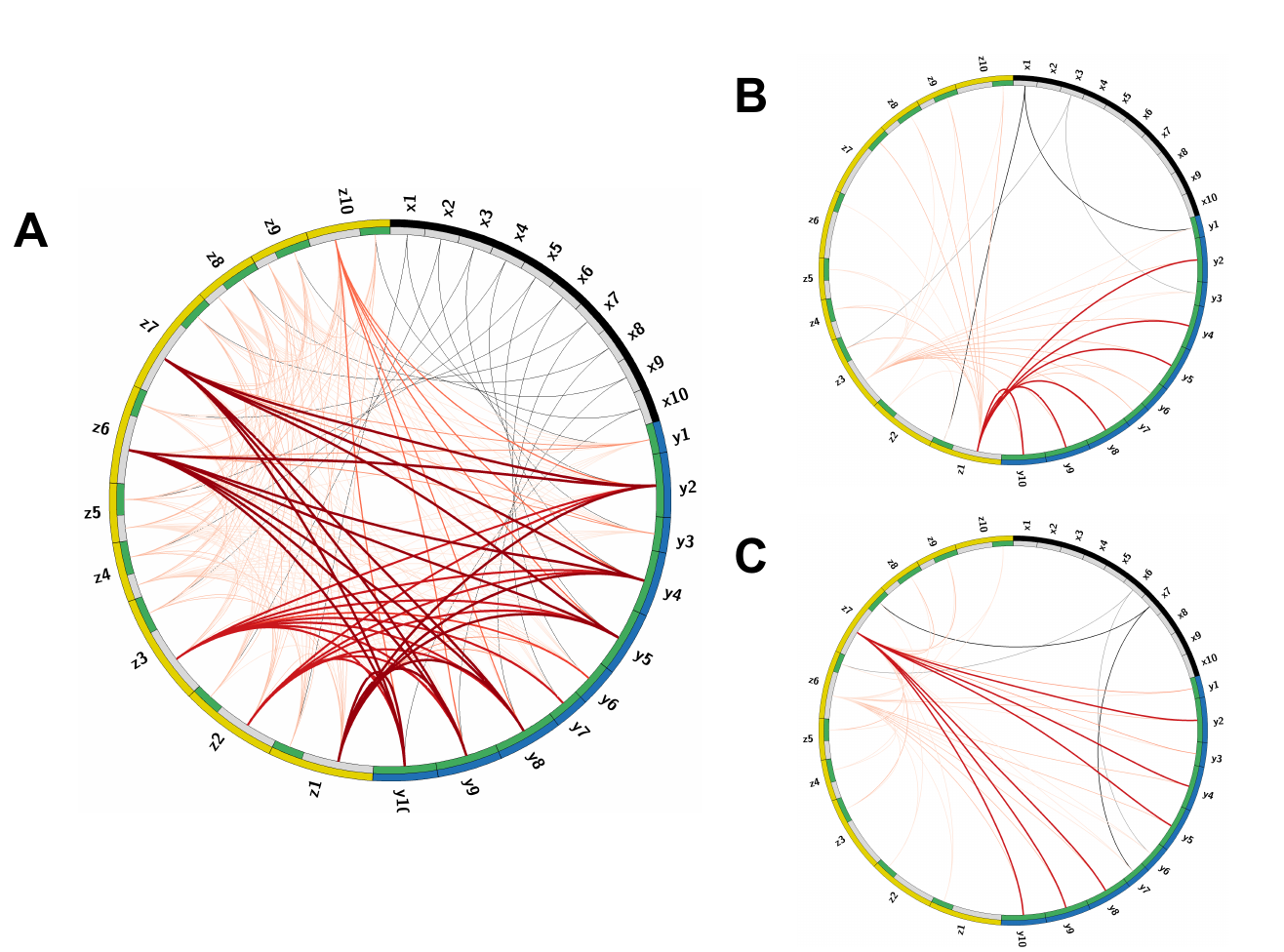}
\caption{ {\bf Visualization of the Connectome.} (A) The connectome (neurons and their
connections) of the reconstructed network depicted in a ring shape. The nodes
on the outer ring correspond to RNs, LNs and PNs and marked by black, yellow
and blue colors respectively. The inner ring splits each node into  ``in"
and ``out" terminals colored by  gray and green colors.  The edges denote
connections between the nodes. 
Excitatory connections are displayed by  gray color and inhibitory by red color. The shades and width  of the edges denote the strength of the connections
(darker and wider curve corresponds to a stronger connection).  (B,C) Activated pathways in the connectome when
 the input is the key: $\vec J_1$ (B) or $\vec J_2$ (C). 
The shades   and width of the ``out" edges in the connectome scaled by the input into the node. When there is no input into the node, its ``out"
edge is not shown.}
\label{fig:2a}
\end{figure*}
\subsection{Recovering the Connectome of an Example Network}

As an illustrative example of the theoretical construct proposed here,
we demonstrate how we establish  the neuronal wiring on a network of 10
neurons of each type:  10 RNs, 10 PNs and 10 LNs
for a total of 30 neurons. 
The network is designed to encode  two input keys into two
output population encoding vectors (codes)  identical to the input keys. The
goal of the calibration is to determine the connectivity matrix $B$
given the matrices $A,C$ and $E$ (Fig.~\ref{fig:1}C).  Specifically,
we choose the  matrices $A$ and $C$ to be identity matrices, i.e.,
each receptor is connected to its  corresponding PN and LN. The matrix
$E$ is set as a random matrix whose elements are drawn from a uniform
distribution with mean $0.25$, i.e., the LNs are randomly connected
between themselves. We then solve an optimization problem, Eq.~(\ref{eq:minprob}),
derived in
the Methods section, to  determine the
elements of the matrix $B$. This is the optimal matrix that supports
such an input-output relation (Fig.~\ref{fig:2}B). The matrices are 
asymmetric, showing that our approach is consistent with experimental
anatomical data.  Moreover, it is fundamentally different than the
Hopfield-type approach that uses symmetry constraint for
optimization~\cite{Reisenman2008,Hopfield1986}.   

The calibration process produces connectivity matrices from which the
connectome of the full network is recovered. To visualize the connectome we use the CIRCOS
package~\cite{Krzywinski2009} ({Fig.~\ref{fig:2a}A) where the network is depicted in a ring shape: FR units are drawn as arcs
on the ring's perimeter and the connections  are the links between the
arcs.  The    connectome structure allows us to observe that 
indeed  the  remainder PNs, labeled as ($y2,y4-y6,y9,y10$), have stronger input
inhibitory connections (dark bold red curves) than the PNs that
participate in the output codes, labeled as ($y1,y3,y6,y7$). We
further observe that these strong connections are  output connections
of LNs, activated by RNs participating in one of the keys, 
labeled as ($x1,x3,x6,x7$). This confirms that the strong inhibition of the
remainder PNs is activated only when there is enough input from RNs
participating in the keys. In addition, each input key activates the
suppression of the other key, though less strongly than the
suppression of the remainder. This is expected from the calibration
matrix $W$ specification (see caption of Fig.~\ref{fig:2},
and the definition of $W$ in Eq.~(\ref{eq:Wdef}). The random connections between the LNs, defined by the
connectivity matrix $E$, are seen in the graph as edges marked by
light red color.        

Once the connections are determined, the
deterministic dynamics of the  calibrated connectome defined in Eqs.
(\ref{eq:FRsys1})-(\ref{eq:FRsys3}) can be explored computationally in
order to verify that  the calibration gives the  desired
low-dimensional dynamics.  In Fig.~\ref{fig:2a}B,C we depict the
\emph{active pathways} in the connectome, i.e., the pathways activated by the input keys. We demonstrate
that for the input $\vec J_1$ (Fig.~\ref{fig:2a}B) four excitatory  edges
are activated in the connectome, where the edges from $x1$ are
stronger  than from $x3$ as expected. These edges excite LNs that
activate inhibitory pathways to PNs.  The strongest  inhibitory
pathway is invoked by $z1$ that suppresses strongly all remainder PNs.
There is also relatively strong suppression of the PNs that
participate in the input key $\vec J_2$ and  very weak suppression of PNs
that should be activated when the input is $\vec J_1$. For the input key
$\vec J_2$ (Fig.~\ref{fig:2a}C) the remainder is strongly suppressed again,
but by a different LN ($z7$).  Moreover, the suppression of neurons
that should respond to $\vec J_1$ is stronger than that of $\vec J_2$ , i.e. the suppression is switched as expected to support $\vec J_2$ instead of $\vec J_1$.
  
From the structure of the effective  connectome, we can conclude
that it indeed  produces the expected \textit{low-dimensional} dynamics. 
Further verification is shown in Fig.~\ref{fig:2}C where the dynamics of
the full network are exactly the dynamics of the prescribed projected
low dimensional system, Eq.~(\ref{eq:projeq}).  When $\vec J_1$ is the input, Fig.~\ref{fig:2}C (left), all
trajectories are attracted to a unique stable fixed point on the
vertical axis, and when the input is $\vec J_2$, Fig.~\ref{fig:2}C (right),
the trajectory is attracted to the unique stable fixed point on the
horizontal axis.
%
%
\begin{figure*}[t]
\centering
\includegraphics[width=0.9\textwidth]{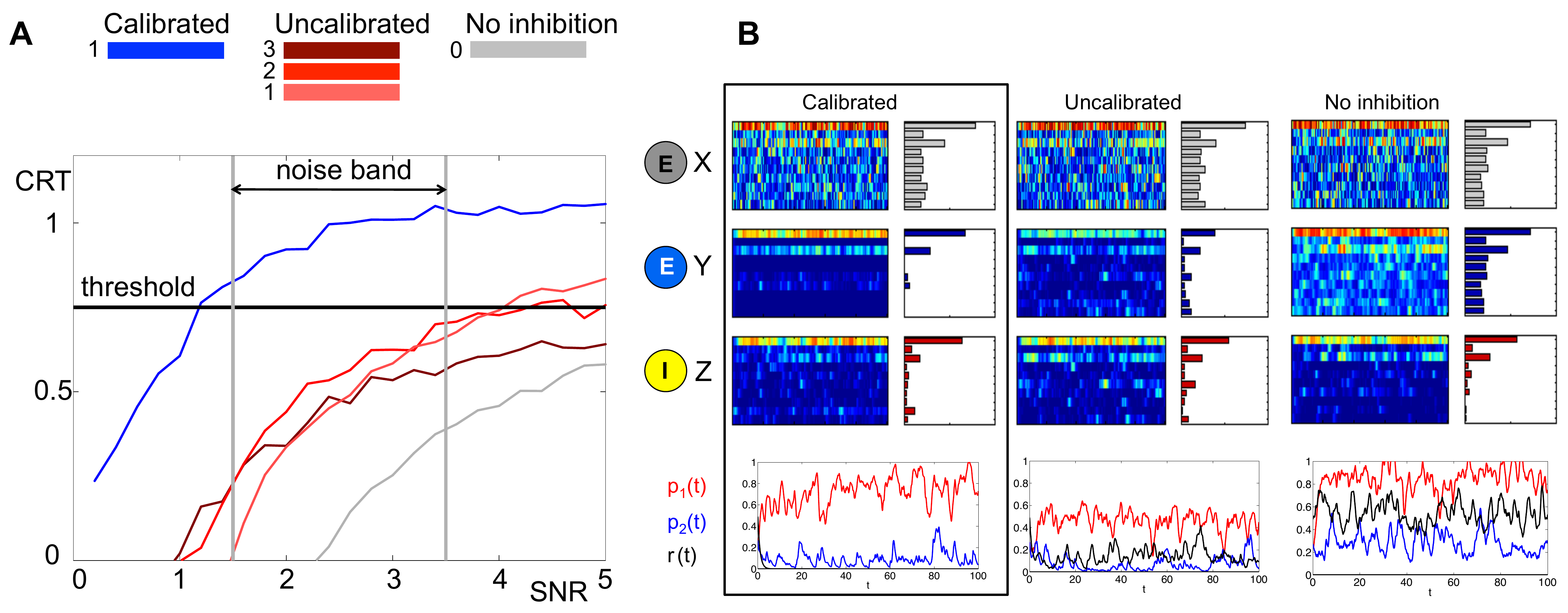}
\caption{{\bf Comparison of stochastic dynamics of an example network
calibrated as in Fig.~\ref{fig:2}B,C.} (A) Contrast (CRT averaged over 10 runs) vs. SNR for different choices of the connectivity matrix $B$: calibrated by~(\ref{eq:minprob})
(blue),
 elements drawn from a
random uniform distribution  with $m\times0.5$ mean:
$m=1$(pink),  $m=2$ (red) and $m=3$ (brown), connections are blocked $B\equiv
0$ (gray). (B)~Dynamics of a network with noisy input $\vec J_1$  (signal-to-noise ratio SNR=3) for different choices of the connectivity
matrix $B$ (from left to right): calibrated by (\ref{eq:minprob}), elements drawn from
random uniform distribution with mean $0.5$, connections are blocked $B\equiv\ 0$. The elements
of the matrix $E$ are drawn once from a uniform distribution and fixed. The color raster plots show the FR dynamics of the
neurons in  X (RNs), Y (PNs) and  Z (LNs) classes (blue:low FR, red:high FR). The  bar plots on the
right side of each raster plot show the average
FRs over the whole evolution. The bottom plots show   the projection of
Y neurons onto the patterns $\vec o^{P}_1,\vec o^{P}_2$ and $\vec o^{R}$ corresponding
to $p_1(t),~p_2(t)$ and $r(t)$ and depicted with \textit{red, blue} and \textit{black} colors
respectively. }
\label{fig:3}
\end{figure*} 
    
\subsection{Noisy Inputs}
Input into the AL varies significantly as a function of time due to
environmental effects, thus producing low signal-to-noise ratio input
signals to the AL.  We can use the example network as a prototype
system to study the stochastic dynamics of such networks and the
implications on the calibration proposed here.  To simulate
noisy inputs, we define the input as $\vec J = \alpha \vec J_k+\sigma \eta(t)$
and define the signal-to-noise ratio (SNR) as $\alpha/\sigma$.  Our
objective is to verify that for different SNR ratios,  the performance
of the network produces the correct population encoding vector.  As
demonstrated in experimental studies of olfaction and other sensory
systems, the underlying neural networks for signal processing, e.g.,
the AL in olfaction,  produce contrast enhancement.  To quantify the
contrast enhacement, we introduce the measure, contrast over time
(CRT),  for a noisy input key $\vec J_k$, defined as
$CRT_k=p_k(t)-\sum^{l}_{j=1,j\neq k}p_j(t)-r(t)$.  This describes the
difference between the $k$-th population encoding vector, $p_k(t)$, and the
summed dynamics over all other population encoding vectors, $p_j(t)$, and the
remainder, $r(t)$.

Intuitively, the measure will be larger when there is a better
separation between the correct input and all other possible inputs. We
investigate the average CRT over time vs. SNR in Fig.~\ref{fig:3}
(left) for  three different network structures where the matrix  $B$
is \emph{calibrated}, \emph{uncalibrated} (random  with different
magnitudes) and has \emph{no inhibiton} (all zeros).  It can be 
clearly observed that the calibrated network achieves the best CRT out
of all other network wirings. The calibrated network exhibits a
$1.5-4$ fold increase in CRT values in comparison to its corresponding
uncalibrated networks, and a  $10$-fold increase over the case where
there is no inhibition.   In particular, network calibration is
important at low SNR rations ($1.5$ to $3.5$), which is the expected
noise band in the actual
environment~\cite{Bhandawat2007,Riffell2008,Riffell2009a}.  
Otherwise, the correct population  encoding vector cannot be separated
from the background noise, Fig. \ref{fig:3}A.  Indeed, only the
calibrated CRT curve is able to cross the 0.75 CRT threshold
(approximately 75\% of separation) in that SNR band.  By varying the
amplitude of the uncalibrated connections we illustrate that  the
amplitude of the elements in B do not necessarily improve the CRT.
When the amplitude is low (see gray curve for $0$ amplitude) the
performance is poor because the activity is noisy. Incrementally
increasing the amplitude improves the performance such that it is able
to cross the CRT threshold when SNR exceeds 4 (red curves). However,
for a calibrated network the crossing of the threshold happens for
much lower values of SNR. Remarkably, even for SNR lower than 1 (where
noise prevails over the signal) the calibrated CRT curve (blue) is
already crossing the threshold.  Additional increase in the amplitude
of  the inhibitory connections will inhibit both noise and the signal,
and we indeed observe that the CRT curve (brown) drops lower than the
lowest amplitude curves and does not cross threshold in the 0-5 SNR
band.

To understand the contrast enhancement more intuitively,   we show in Fig.~\ref{fig:3}B the dynamics for SNR=3.
  At this SNR, the dynamics of RNs and LNs are very similar for all
network wirings. The dynamics of RNs are noisy, making it  very
difficult to recover the input key from the data. The dynamics of LNs
are cleaner, but still do not have a clear signature of the input key
signal.  In particular the ratio between the two elements of the key,
neurons $1$ and $3$, is incorrect. The dynamics of the PNs, however,
are very different for the three choices of  network wirings. In the
calibrated network the dynamics of PNs are more distinguishable
relative to  other networks. Indeed, both FRs over time and average
FRs  indicate that the output signal is the closest to the population
encoding vector  $o^P_1$ corresponding to the input key $\vec J_1$ 
(the CRT value is around 1). For uncalibrated or no lateral inhibition
wirings, such a clear signature cannot be detected. Indeed the CRT measure
for uncalibrated network is 0.55 and for no inhibition network is 0.2.
These dynamics reflect  the underlying network structure.  Thus for
the uncalibrated network, all population encoding vectors are being
inhibited proportionally so that detection of the correct output  both
from the averaged FRs histogram and/or the projected dynamics is more
difficult.  Similar behavior, but even more noisy, occurs when there
is no lateral inhibition. 

We also compared the calibrated and uncalibrated
wirings obtained from data (stimuli C and D) by adding noise of $\sigma=0.3$ (SNR=3) 
to the models and computing the CRT trajectory over time for each simulation (5 simulations per wiring), see Fig. Sup. 1. 
Indeed, the CRT trajectories produced by the calibrated model cross  the correct threshold while 
trajectories produced by the random model do not cross it. In addition, when the stimulus is off the trajectories produced by 
randomly wired model are very sensitive to noise to the extent that they can cross the wrong threshold.

%
%

\begin{figure}[!t]
\centering
\includegraphics[width=0.5\textwidth]{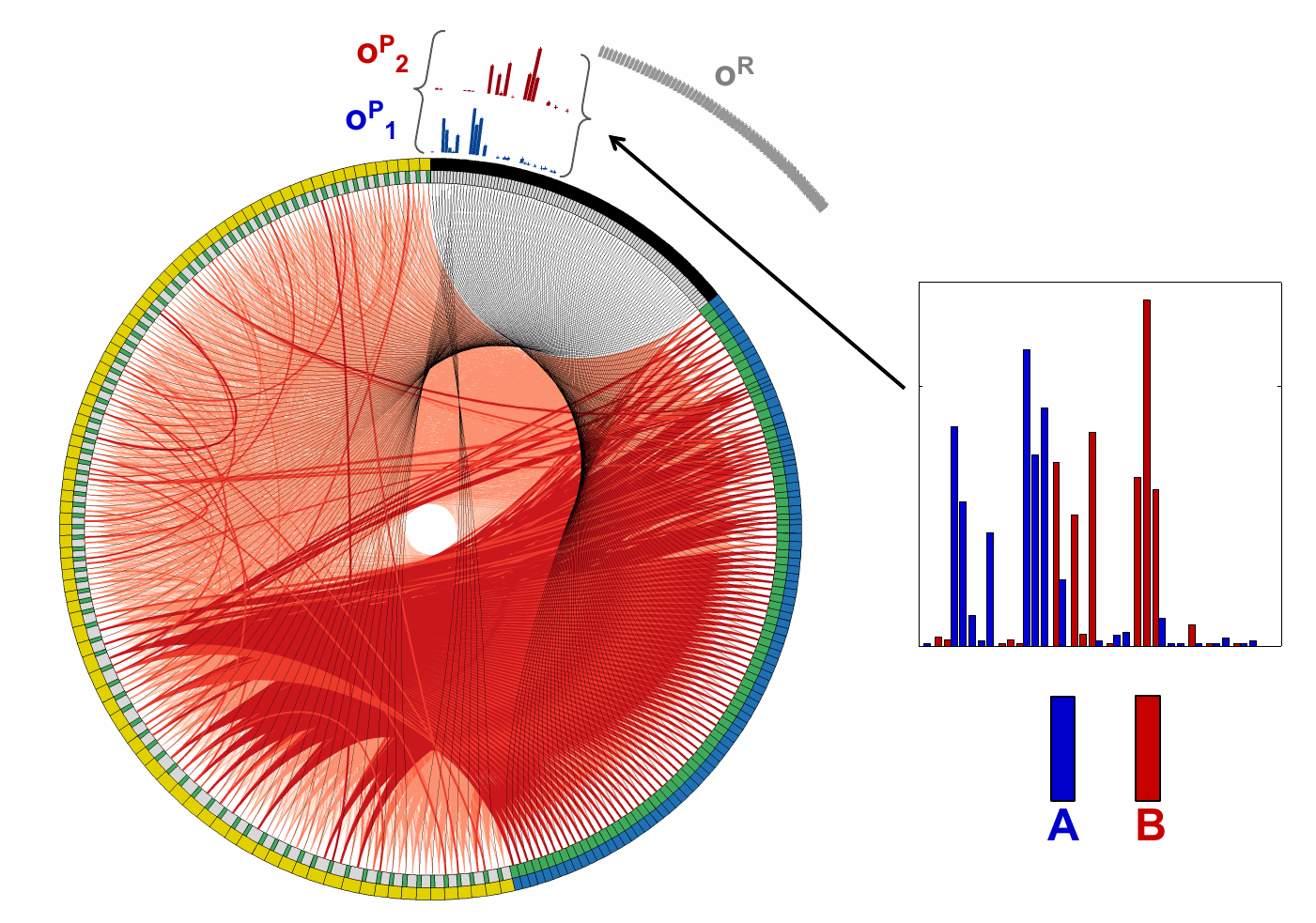}
\caption{{\bf Library elements and the reconstructed connectome produced
from experimental data.}  The connectome of a network of 97 neurons
of each type (291 in total), obtained from solving the minimization problem
in~(\ref{eq:minprob}) with the matrices $A,C$ chosen as   identity
matrices and $E$ as a positive random matrix with elements drawn from a
normal distribution, i.e., $e_{ij}\sim|N(0,0.02)|$.  The connectome is
visualized in a ring shape with similar choice of colors as in Fig.~\ref{fig:2a}A.
The orthogonal library matrix, $O$, used in the minimization, was obtained
from the electrophysiologically  recorded data of two odorants A and B
as described
in  \S\S Construction of the Data-Driven Model in Results. The  three library
vectors $\vec o^P_1$, $\vec o^P_2$ and $\vec o^R$ in
$O$, correspond to the odorants A (blue), B (red)  and the remainder (gray) respectively, are
depicted as bar-plots above the connectome at the locations that correspond to RNs that
evoke each of the vectors. 
To the right of the connectome, the   bar-plots of the  vectors $\vec o^P_1$ and $\vec o^P_2$ are enlarged.   The threshold
value, $\tau$, used to construct the library vectors, is $\tau=0.07$.}
\label{fig:4}
\end{figure}
\subsection{Construction of the Data-Driven Model}
 We proceed and construct  a dynamical model using the experimental data. In the first series of experiments
 we recorded  from 130 PNs
 that were stimulated with two odorants: ``A" (BAL-Bombykal),
 ``B" (MYR-$\beta$-Myrcene). These stimuli are behaviorally effective odorants: odorant A is a component of the moth sex
 pheromone, and odorant B is a flower scent component. These odorants excite distinct glomeruli in the AL and require
 a minimal orthogonalization of the library.  Therefore, we chose them to validate the
 data-driven model construction part of our approach. Another reason for the choice is that they are (negatively) correlated with each other --
 when a particular stimulus is on, PNs associated with it are excited
 while those associated with the other stimulus are inhibited (see Fig 1 Sup). This suggests that these regions inhibit each other via lateral inhibition.
 We also recorded from 70 PNs with two related stimuli: ``C" (BEA-benzaldehyde), ``D" (BOL-benzyl alcohol) that excite PNs in overlapping glomeruli.  Both odorants are dominant in floral scents related in chemical structure as oxygenated aromatic volatiles.
 
 The odorants are presented to the preparation at a realistic time interval (200
 msec) repeatedly for five stimulations separated by long intervals of
 no input. For more information regarding the experimental setup and
 procedures see the subsection `Electrophysiological preparation and
 stimulation' in the Methods section. The data is available in the supplementary
material.  With the  spike trains of each PN 
 we have computed the time series of the  instantaneous FR (iFR)
 averaged over the 5  trials of odor introduction. Sampling the iFR at
 a specific time after the beginning of the odor introduction (at 150 msec) or performing a PCA reduction
 and taking the first dominant mode, we obtained a histogram of iFRs
 for  the neurons for each of the odors. The neurons with substantial
 difference in iFR in response to the two odorants were assigned as
 selective neurons (37 neurons for A,B and 30 neurons for C,D).  Those with low iFR were assigned as
 remainder neurons (60 neurons for A,B and 40 neurons for C,D).   The remaining neurons that
 exhibited high iFR were not included in the calibration (33 neurons for A,B)
 since there was not enough data to calibrate the inhibitory
 connections to them. 

 Application of the orthogonalization procedure, defined in the Methods
section, for the  97 neurons for A,B and 70 neurons for C,D
 resulted in the two population encoding vectors: $\vec o_1^{P}$ for A
 (blue) and $\vec o_2^{P}$ for B (red) as shown in Fig.~\ref{fig:4} (vectors for C,D 
 are similar so were not shown). For A,B the required orthogonalization is minimal, 
 while for C,D it is significant as shown in Fig 2 Sup.
 
 This allows for the reconstruction of the \textit{connectome} of the  AL
 network in a similar fashion to the example network.  Here the matrix
 $E$ is taken as a random normal matrix and the matrix $B$ is
 calibrated.   The full network consists of the three populations
 (PNs, LNs, RNs) of 97 neurons (291 neurons in total),   where in each
 population we depict (in the clockwise direction) the selective
 neurons followed by the remainder neurons (Fig.~\ref{fig:4}).
 Although many connections exist, the ring shaped  visualization
 demonstrates qualitatively the main features of the connectome:  (i)
 the suppression of the selective neurons seems to be nonuniform and
 sparse while  (ii) the inhibition of remainder neurons is uniform and
 dense. These biophysical wirings are consistent with the derived
 population encoding vectors that are nonuniform.

\subsection{The Dynamics of  Population Encoding Vectors}
The orthogonality of the population encoding vectors, $\vec o_1^{P}$ and $\vec
o_2^{P}$,  allows us to construct a two dimensional space, called the \emph{odor  space}.  
We use it to project the iFR time series obtained from either the data or the calibrated model. The data projection is used to assess whether
the experimental dynamics are consistent with the underlying 
dynamical mechanism in the construction of the model.  Specifically,
there is a single, stable fixed point in each encoding vector
direction.    Figure~\ref{fig:5op1} 
shows the projection dynamics
(gray) of five experimental trials along with the average trajectory
over the trials (black).  Before the input is applied,  the projected
trajectory hovers around vicinity of the origin due to noise
fluctuations. When the input is applied, the trajectory begins an
excursion toward the stable fixed point and when the input is off, the
trajectory returns to within the vicinity of the origin. The input of
odor ``A", ``C" corresponds to trajectories whose fixed point lies on the
vertical axis while odor ``B", ``D" trajectories evolve trajectories towards a
fixed point on the horizontal axis.

%
%

\begin{figure*}[!t]
\centering
\includegraphics[width=1\textwidth]{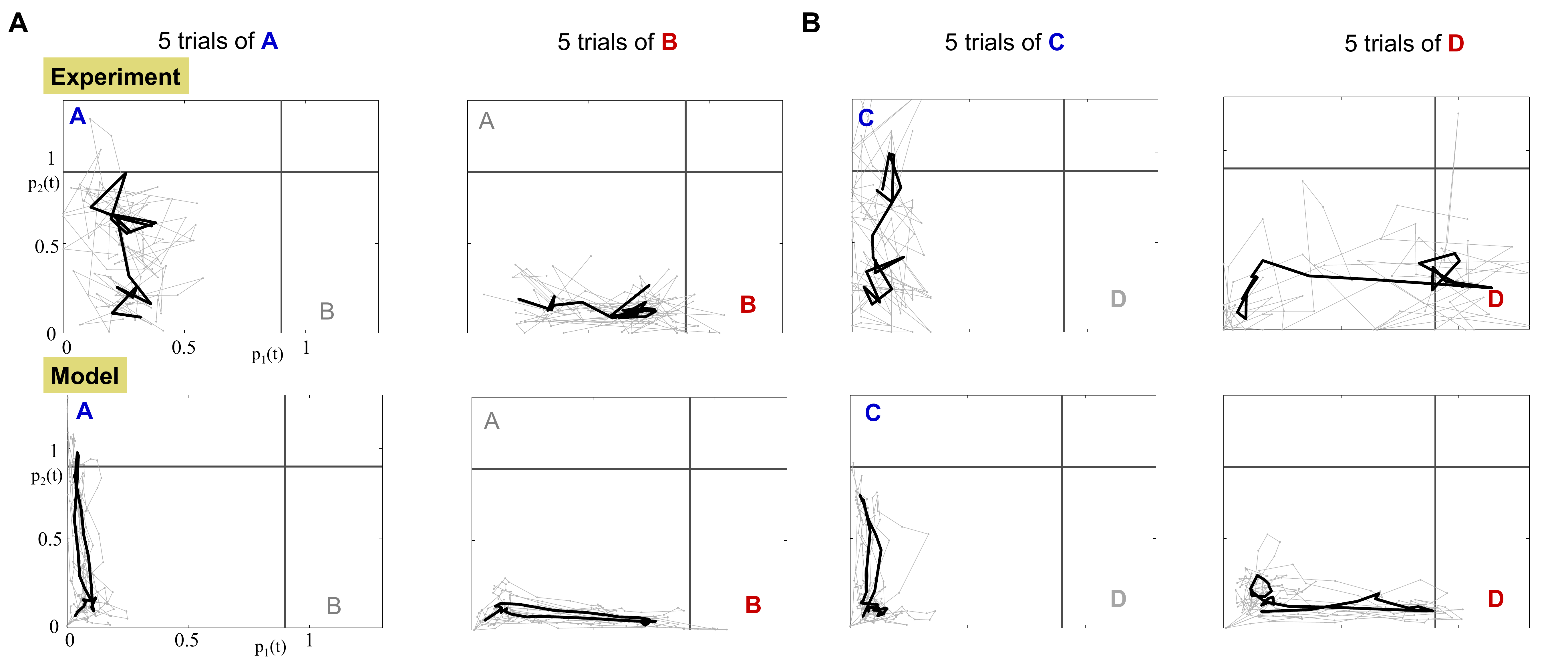}
\caption{
{\bf Projection of the experimental and model dynamics onto orthogonal
odor space.}  Top row: projected experimental dynamics of  PNs.
Bottom row: projected dynamics of the calibrated model. Each column shows
5 distinct trials per stimulus; Panel A: for A,B stimuli; Panel B: for C,D stimuli.  Gray trajectories
are 5 distinct trials of the application of the odor. The starting and  ending
points of the plotted trajectories are 100 msec and 600 msec, respectively, after the beginning of the trial. The black bold trajectory
is the averaged trajectory over the trials.  }
\label{fig:5op1}
\end{figure*}


Data projections from both the experimental data and the calibrated
model, Fig. \ref{fig:5op1} and supplementary videos, clearly demonstrate that different odorant
inputs correspond to different orthogonal fixed points in the
projection space. Furthermore, trajectories appear   noisy while reaching
the fixed point whereupon they remain static for a while until the input is
stopped and then trajectory returns to the origin.

\begin{figure}[!t]
\centering
\includegraphics[width=0.5\textwidth]{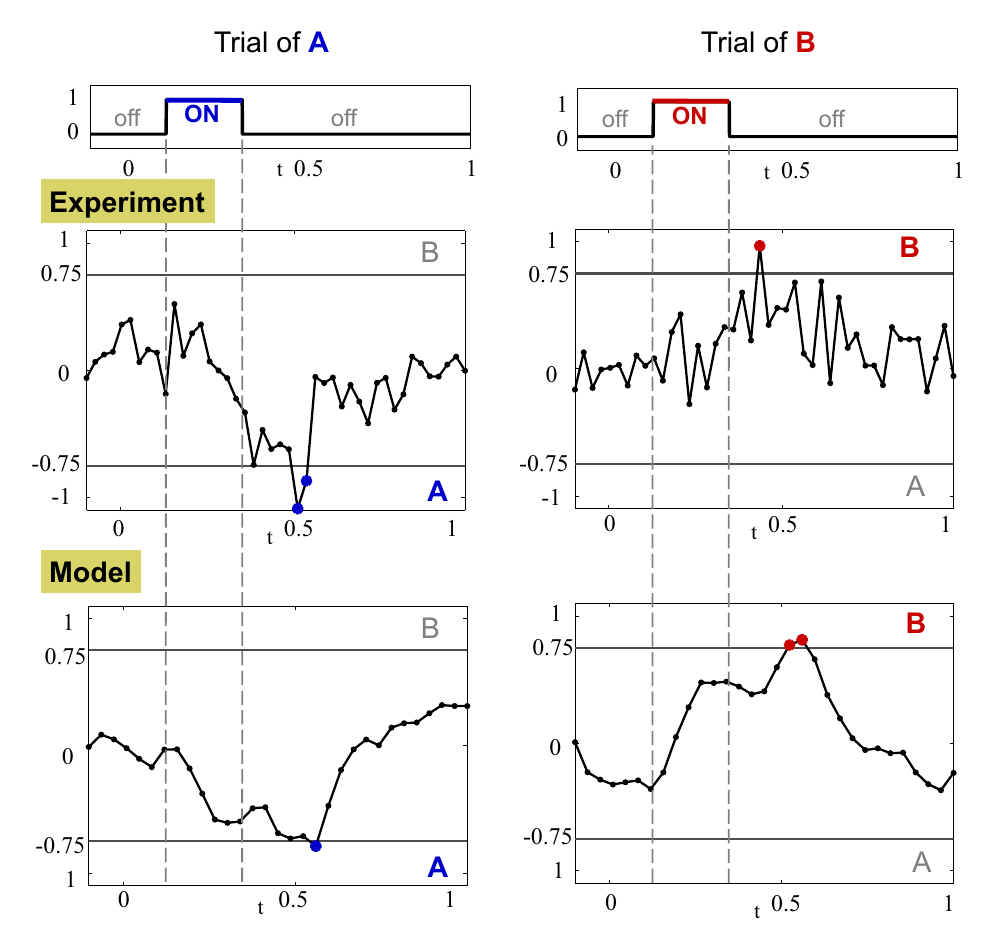}
\caption{
{\bf CRT dynamics demonstrates the decision making associated
with odor detection.} Left and right
 columns show
the dynamics of $CRT_{A/B}$ measure for one trial of  stimulus A and B
respectively. The CRT measure computed for stimuli A and B is: $CRT_{A/B}
=p_B(t)-p_A(t) \pm r(t)$.  Top row: the amplitude of the stimulus input over time. Middle and bottom rows: CRT
dynamics, in the experiment  and in the calibrated model respectively.
  The threshold decision lines are plotted with gray color at -0.75 (A) and 0.75 (B). When the CRT curve (black  line with dots corresponding to
 data points) crosses the threshold,
the data points  are marked with appropriate color of their dots: blue for stimulus A and red for stimulus B. }
\label{fig:5op2}
\end{figure}

\subsection{Decision Making}
In the experiments described here, the presentation of a stimulus odor
occurs for an extremely short period  of time (approximately 200~ms).
Such inputs correspond to realistic stimulus for which  the moth is
flying and sampling odors in a turbulent environment. 
Thus once we have characterized the
dynamics of each short trial, we  aim to deduce the  
dynamical mechanism that corresponds to  a sequence of short bursts of
inputs and its benefits  on odor detection and selection.

To formulate the decision making process, we analyze the dynamics of a
trajectory toward the orthogonal fixed point when the stimulus is
introduced as demonstrated in Fig.~\ref{fig:5op1}. 
The orthogonality of the fixed points allows us to construct 
\emph{threshold lines} for determining  odor detection.   The gray horizontal and vertical lines in
Fig.~\ref{fig:5op1} represent the threshold for the detection of odor
A  and odor B respectively. Application of a single odorant ensures
that the dynamical trajectory crosses  only a \textit{single} threshold line on
its way to its corresponding fixed point. Experiments show that it
spends only a small amount of time near the fixed point (approximately
100 msec) before returning back to the origin.

While it is difficult to measure the convergence rate of the
trajectory to the fixed point, it is straightforward to detect a
crossing of the threshold line. Indeed, a common hypothesis in decision making associates 
crossing of a threshold in neuronal activity as equivalent  to making a
decision~\cite{Bogacz2006,WongWang06}.   The crossing of the threshold line  in our case can be
captured most effectively by computing the CRT measure, shown in
Fig.~\ref{fig:5op2}, per each trajectory of each odorant. Results from
this analysis demonstrate that after the input is introduced, the CRT
curve tends toward one of the decision thresholds, passes it and then
returns back to the region where no clear contrast exists between
odorants.  
Thereby, passing of the threshold creates
an evidence toward one of the odorants. Integration of such evidence over several trials can produce a
significant bias towards a specific odorant. When enough
crossings from trial to trial occur, strong evidence is accumulated to
accurately determine an odorant.

From simulations we observe that there is a clear advantage in repetitive introduction of the input in
short bursts rather than a single long input.  Long input, when noisy,
 creates a corresponding noisy trajectory that typically  crosses the threshold line a
single time and then wanders around the fixed point so that the decision is
based only on one evidence. Other measures
such as the time the trajectory spent near the fixed point are typically non-robust
when noisy dynamics are considered.   In contrast, repetitive inputs generate a mechanism that
allows for better integration of evidence since for each trial it is
enough to just cross the threshold once given by a  simple measure
like the CRT.  Such a mechanism thus provides a rapid, robust  approach
to odor detection. 

\section{Discussion}
The  discovery that neural responses in the antennal lobe (AL), olfactory primary processing unit in insects, are encoded into spatio-temporal neural
codes suggests that the dynamics of odor recognition and perception are fundamentally
a low-dimensional processing phenomenon.  Characterizing the dynamical mechanism responsible for such 
functionality is a critical step towards understanding of 
the  neurobiological networks involved in odor perception~\cite{Rabinovich2008}.  We modeled the  dynamics of the AL by developing a data-driven network model for  the  neural
codes that appear in the recorded data from AL projection neurons. 
The methodology introduced here associates a population encoding vector
with each of the neural codes and  derives a low-dimensional model for
the  dynamics of these vectors.
The low-dimensional model is then used  to calibrate the full computational
model of the AL. Effectively, our analysis produces a high-dimensional 
network of neurons tuned to encode scents into low-dimensional activity
patters, i.e., the neural codes.

Comparison between the model and experiment shows strong quantitative 
agreement and supports the underlying theoretical concept of the dynamical dimension reduction for
robust encoding.  Moreover, it allows us to construct an optimal projection space for 
understanding odor recognition, characterize the role of lateral inhibition, and propose a scheme for the integration of evidence for dynamic 
decision making.

\subsection{High-Dimensional Neural Network is Tuned to Exhibit Low-Dimensional Dynamics }

Our primary contribution is the introduction of a new methodology
that combines dimensionality reduction of dynamical systems with experimental data in order
to achieve a reliable computational model, that highlights the exploitation of low-dimensional
encoding in the AL. To our knowledge, this is the first successful model that
combines such a data-driven methodology in conjunction with dynamical equations of FR activity.

The methodology is divided into two stages.  The first stage is implicit, where
we define a library of population (encoding) vectors and project the dynamic neuronal network 
onto these vectors~\cite{Sirovich1996,eshamee11}.  
The outcome is then restricted so that the system possesses stable orthogonal
fixed points, where each orthogonal direction is associated with a different population encoding vector. 
This restriction determines a mapping from the high- to low-dimensional system. 
In the second stage of the modeling,  the constructed mapping is used in conjunction with 
experimental recordings in order to both determine the population encoding vectors and 
reconstruct the AL connectome associated with each odor in the library
(see Fig.~\ref{fig:3}).
The construct is consistent with AL experimentally described functionality:
while it is a high-dimensional neural network  consisting of thousands of neurons, it
appears to be tuned to exhibit  low-dimensional coding dynamics.

With this theoretical framework established,  we  are able to suggest  answers to
key questions in the behavior of the AL.  
One of primary importance is identifying the optimal network design that maximizes contrast enhancement  and reproduces the observed AL functionality.  The model shows that the optimal design can be
constructed by  tuning  the lateral-inhibition so
that the patterns of FR activity are made robust \cite{Rabinovich2008}.
In particular, we show  that asymmetric, nonlocal design of connections in a network of neurons
can lead to such low dimensional robust functionality. This mechanism
is distinct from the mechanisms in other sensory systems, e.g., vision,  and thereby demonstrates that common wiring strategies across different sensory systems is not necessary.   

Furthermore, we demonstrate that in a noisy environment, network tuning is \emph{necessary}  for
robust detection of an odor, even when  input keys and output codes are identical. We show that lateral-inhibition,
that has been tuned,  shapes the noisy input into reliable and repeatable trajectories, while  inhibition that was not tuned produces 
 noisy and unreliable trajectories. This phenomenon is experimentally observed and  described as  contrast enhancement.  
 Furthermore, our work suggests that absence of lateral inhibition will result in noisier responses and scattered trajectories in the odor space. 
These prediction can be verified by pharmacological treatment of the AL with GABA antagonists that block inhibition.

\subsection{Projection Space for Odor Detection  }

In previous studies, a three dimensional  projection space (\emph{odor
space}) was  constructed using PCA based dimension reduction.   Projections
of  distinct
odorant trajectories  onto this low-dimensional space appeared  to be   well separated from
each other~\cite  {Laurent2002,Mazor2005}.  Moreover, for each odor
there was an associated fixed point that  was  separated from all
other odor fixed points.  The construction demonstrated that odorants 
can be classified into distinct groups and suggested that odor
detection may be accomplished solely from recordings and projection
onto the odor space.

The odor space is the backbone of our model as well. There
are key differences, however, in the construction of our underlying odor space.   
Specifically, we treat the data differently by dividing
the population of PNs into remainder  and population encoding vectors
so that we achieve a  model representing the dynamics of the
spatio-temporal FR patterns rather than single neurons.   Such a
viewpoint of the data is useful since it constructs an odor space
(phase space of a dynamical system)  with meaningful axes, i.e., our dimensionality
reduction gives an orthogonal basis as the ideal projection space for odor encoding. 
Specifically, each axis corresponds  to individual odorant or the
remainder (Fig. \ref{fig:2}C). As a result,  the odor space
provides an easy means for characterizing odor recognition since a
given trajectory can be  simply defined.   Indeed, comparison of the trajectories in the model  and the
experiments,   Fig.~\ref{fig:5op1},  confirms that the dynamical
mechanism upon which the model was constructed (stable orthogonal
fixed points) is  the dynamical mechanism associated with odor
detection.

\subsection{Decision Making as a Robust Mechanism for Odor Perception}

The timescales of realistic inputs indicate that odor detection occurs
relatively fast and usually requires repetitive (over several trials) exposure to the same
odor \cite{Mainland2006,Koehl2001}. Some animals use sniffing or other mechanisms to achieve fast repetition 
of similar input into the olfactory system \cite{Mazurek2003,Vickers1994,Mafra-Neto1994,Riffell2009a}. Furthermore,  there exists
experimental evidence that    shows that for a longer stimulus duration (a few seconds), initial  sharp  response of PNs is followed by more intermittent one~\cite{Christensen1988,Marion-Poll1992}. These results
suggest that the optimal strategy for scent recognition is employed by sampling
the stimulus.   Our analysis  suggests that indeed  based on the
dynamics of the AL the exposure to multiple, short-time bursts of
odor can be formulated as a decision making process. More precisely, we are able to prescribe an algorithm,
 possibly  evoked by higher centers in the brain such as the mushroom body,  that polls the dynamics of the AL in order to make a decision. 

Examination of the projections of iFR data produced in both theory and experiment indicates that in each 
short trial, the most plausible dynamical response is an excursion in odor space along a trajectory attracted toward 
an orthogonal fixed point. In fact, the orthogonality of the fixed points  allows for an optimal separation of trajectories
for different odors.  Due to the short timescale of the odor burst, the trajectory does not necessarily converge to the
fixed point.  Rather, it only approaches its vicinity (Fig.~\ref{fig:5op1}
and supplementary videos).  In effect, it crosses the threshold line of an odorant while staying away from crossing thresholds of other odorants, see the horizontal and
vertical lines in Fig.~\ref{fig:5op1} and the trajectories
that cross them. Indeed, a common hypothesis in decision making is that the decision
is made when neuronal activity crosses a threshold~\cite{Bogacz2006}. 
Tracking trajectories that cross decision thresholds
is accomplished by defining a linear contrast measure over time as we demonstrate in  Fig.~\ref{fig:5op2}.
Repetition of the same odorant stimulus permits robustness of the algorithm. With each threshold
crossing, evidence is integrated towards a specific odorant stimulus. After each trial,
a probability distribution is updated until there is a high probability that supports a specific odorant stimulus. 
This indicates that enough evidence was integrated toward one of the odorants, and thus leads to a 
decision/perception for the odor, which is followed by a behavioral response corresponding to that odorant. 

The proposed algorithm is scalable and can be used for the perception of
complex odors, i.e.,  a mixture of odorants.
If the  odorants in the compound
are of similar significance and  strength, then  the trajectories in the odor space may cross
several thresholds of distinct odorants each time that a stimulus is applied. Repeating the 
application of the same stimulus, eventually   will lead to reconstruction of a uniform probability distribution indicative
of the distribution of odorants in the mixture.  Note that to obtain a reliable probability
distribution the  process may require many repetitions than in the detection of a single odorant. When the mixture includes nonequal ratios
of odors,
the  mechanism proposed here suggests that  odorants
will compete with each other via lateral inhibitory connections.
Population encoding vectors that correspond to  more dominant odorants may inhibit  other 
population encoding vectors  strongly and suppress them. Thus   the constructed
probability distribution of crossing the threshold  can be significantly different than the distribution of the odorants in the mixture. Simulations and experiments on complex odors are currently under investigation. Additionally,
the derived model could be used in conjunction with    centrifugal input into the AL to explore the strategies for learning-induced modulation~\cite{Cassenaer2012,Farooqui15062003,Dacks2012}. Such modulatory effects on the encoding properties are work for future studies. The model developed here provides an
efficient platform for performing such studies.


\bibliographystyle{spmpsci}      

\end{document}